\documentclass[
    ,final            
  ]
  {aipproc}

\layoutstyle{6x9}

\begin{document}

\title{Non-perturbative QCD Modeling and Meson Physics}


\author{T. Nguyen}{
  address={Center for Nuclear Research, Department of Physics,  Kent State 
University, Kent OH 44242, USA}
}

\author{N. A. Souchlas}{
  address={Center for Nuclear Research, Department of Physics,  Kent State 
University, Kent OH 44242, USA}
}

\author{P. C. Tandy}{
  address={Center for Nuclear Research, Department of Physics,  Kent State 
University, Kent OH 44242, USA}
}

\begin{abstract}
Using a ladder-rainbow kernel previously established for light quark hadron physics, we explore the extension to masses and electroweak decay constants of ground state pseudoscalar and vector quarkonia and heavy-light mesons in the c- and b-quark regions.   We make a systematic study of the 
effectiveness of a constituent mass concept as a replacement for a heavy quark dressed propagator for such states.   The difference between vector and axial vector current correlators is explored within the same model to provide an estimate of the four quark chiral condensate and the leading distance scale for the onset of non-perturbative phenomena in QCD.   
\end{abstract}

\maketitle


\section{DYSON--SCHWINGER EQUATIONS OF QCD}

A great deal of progress in the QCD modeling of hadron physics has been 
achieved through the use of the ladder-rainbow truncation of the Dyson-Schwinger
equations (DSEs).   The DSEs are the equations of motion of a
quantum field theory.  They form an infinite hierarchy of coupled
integral equations for the Green's functions ($n$-point functions) of
the theory.  Bound states (mesons, baryons) appear as poles in the appropriate
Green's functions, and, e.g., the Bethe-Salpeter bound state equation appears after taking residues in the DSE for the appropriate color singlet vertex. 
For recent reviews on the DSEs and their use in hadron physics, see
Refs.~\cite{Roberts:1994dr,Tandy:1997qf,Alkofer:2000wg,Maris:2003vk}.   

In the Euclidean metric that we use throughout, the DSE for the dressed quark propagator is
\begin{eqnarray}
S(p)^{-1}  &=& Z_2 \, i\,/\!\!\!p + Z_4 \, m(\mu) + Z_1 \int^\Lambda_q \! g^2D_{\mu\nu}(p-q) \, 
        \frac{\lambda^i}{2}\gamma_\mu \, S(q) \, \Gamma^i_\nu(q,p)~,
\label{quarkdse}
\end{eqnarray}
where $D_{\mu\nu}(k)$ is the renormalized dressed-gluon propagator,
$\Gamma^i_\nu(q,p)$ is the renormalized dressed quark-gluon vertex.
The solution of Eq.~(\ref{quarkdse}) is renormalized according to
$S(p)^{-1}=i\gamma\cdot p+m(\mu)$ at a sufficiently large spacelike
$\mu^2$, with $m(\mu)$ the renormalized quark mass at the scale $\mu$.
We use \mbox{$\mu=19\,{\rm GeV}$} for numerical work.   The
renormalization constants $Z_2$ and $Z_4$ depend on the
renormalization point and the regularization mass-scale. 

After taking the residue at the bound state pole of the inhomogeneous DSE for the relevant vertex, 
bound state Bethe-Salpeter equation is 
\begin{eqnarray}
\Gamma^{a\bar{b}}(p_+,p_-) &=& \int^\Lambda_q \! K(p,q;P)S^a(q_+)
                                          \Gamma^{a\bar{b}}(q_+,q_-)S^b(q_-)~~,
\label{bse}
\end{eqnarray}
where $K$ is the renormalized $q\bar{q}$ scattering kernel that is
irreducible with respect to a pair of $q\bar{q}$ lines.  The quark
momenta are \mbox{$q_+ =$} \mbox{$q+\eta P$} and \mbox{$q_- =$} \mbox{$q -$}
\mbox{$(1-\eta) P$} where $\eta$ is the momentum partitioning
parameter.  The choice of $\eta$ is equivalent to a choice of relative
momentum $q$; physical observables should not depend on $\eta$ .
This provides us with a convenient check on numerical methods.   The meson momentum 
satisfies $P^2 = -m^2$.

\begin{figure}
\includegraphics[height=0.3\textheight]{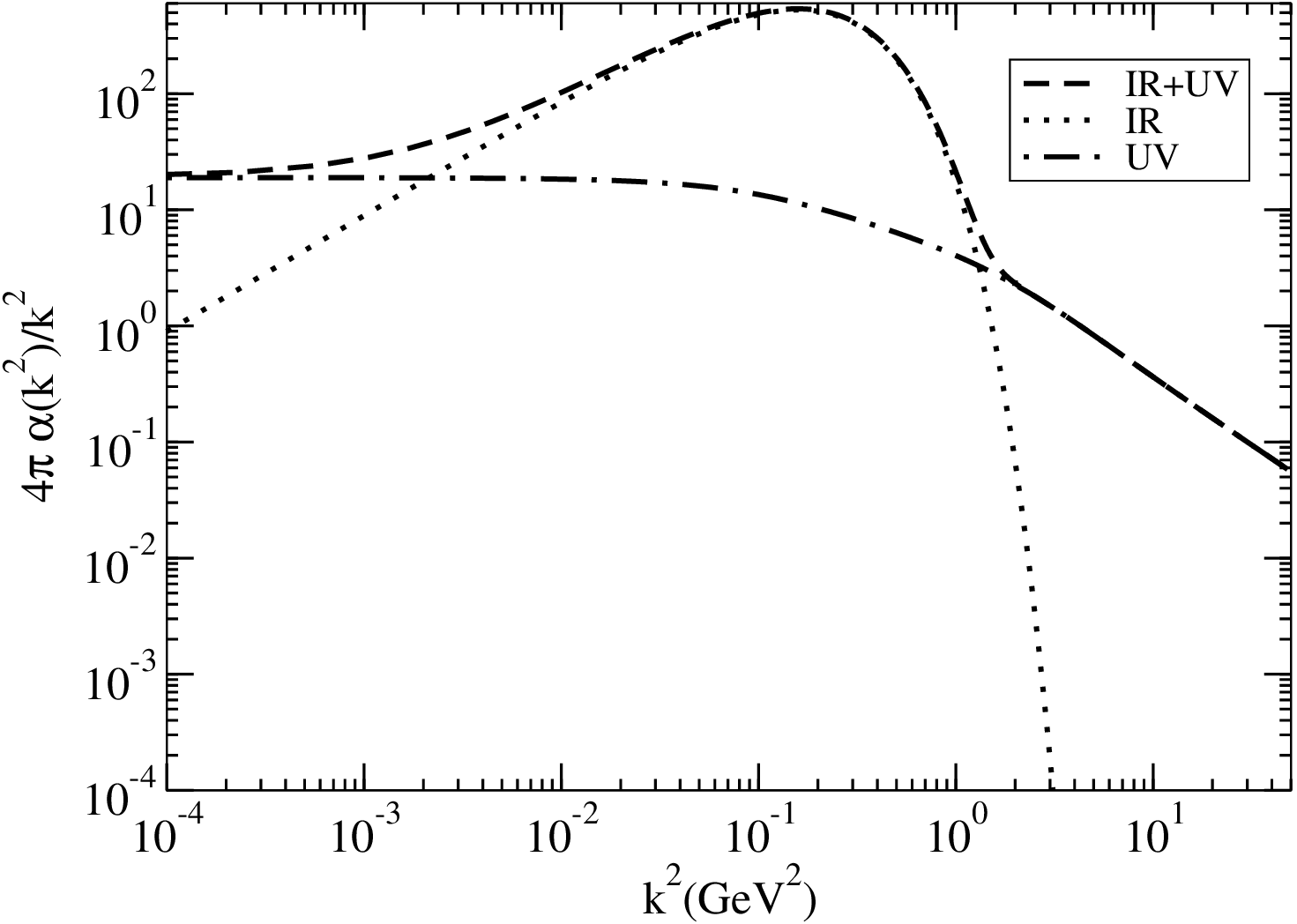}
\caption{The effective ladder-rainbow kernel from the Maris--Tandy [MT]
model~\protect\cite{Maris:1999nt} showing the infrared (IR) and ultraviolet (UV) components.
\label{Fig:MT_kernel} }
\end{figure}

\section{\label{sec:RLMT}Rainbow-Ladder Truncation}

A viable truncation of the infinite set of DSEs should respect
relevant (global) symmetries of QCD such as chiral symmetry, Lorentz
invariance, and renormalization group invariance.  For electromagnetic
interactions and Goldstone bosons we also need to respect color singlet  vector 
and axial vector current conservation.  The rainbow-ladder truncation, which achieves these ends,
replaces the BSE kernel by the (effective) one-gluon exchange
\mbox{$K(p,q;P)  \to -4\pi\,\alpha_{\rm eff}(k^2)\, D_{\mu\nu}^{\rm free}(k)
\textstyle{\frac{\lambda^i}{2}}\gamma_\mu \otimes \textstyle{\frac{\lambda^i}{2}}\gamma_\nu $}
along with the replacement of the DSE kernel for $S(p)$ by
\mbox{$ Z_1 g^2 D_{\mu \nu}(k) \Gamma^i_\nu(q,p) \to 
 4\pi\,\alpha_{\rm eff}(k^2) \, D_{\mu\nu}^{\rm free}(k)\, \gamma_\nu
                                        \textstyle\frac{\lambda^i}{2} $}
where $k=p-q$, and $\alpha_{\rm eff}(k^2)$ is an effective running
coupling.   This truncation is the first term in a systematic
expansion~\cite{Bender:1996bb,Bhagwat:2004hn} of the quark-antiquark scattering kernel
$K$; asymptotically, it reduces to leading-order perturbation theory.
Furthermore, these two truncations are mutually consistent in the
sense that the combination produces color singlet vector and axial-vector vertices
satisfying their respective Ward identities.  In the axial case, this
ensures that the chiral limit ground state pseudoscalar bound states
are the massless Goldstone bosons associated with chiral symmetry
breaking~\cite{Maris:1998hd,Maris:1997tm}.  In the vector case, this
ensures, in combination with impulse approximation, electromagnetic
current conservation~\cite{Roberts:1996hh}.  

\begin{table}
\caption{DSE results~\protect\cite{Maris:1999nt} for pseudoscalar and vector meson masses and electroweak decay constants, together with experimental data~\protect\cite{PDG04}.   Units are GeV except where indicated.   Quantities marked by $\dagger$ are fitted with the indicated current quark masses and the infrared strength parameter of the ladder-rainbow kernel.  \label{Table:model} }

\begin{tabular}{|l|cc|cc|cc|cc|cc|} \hline 
  \multicolumn{2}{|c|}{ }   & \multicolumn{3}{c|}{$m^{u=d}_{\mu=1 {\rm GeV}}$}  &  \multicolumn{3}{c|}{$m^{s}_{\mu=1 {\rm GeV}}$}  &   \multicolumn{3}{c|}{- $\langle \bar q q \rangle^0_{\mu=1 {\rm GeV}}$}    \\   \hline
 \multicolumn{2}{|c|}{ expt }   &  \multicolumn{3}{c|}{ 3 - 6 MeV}   &  \multicolumn{3}{c|}{  80 - 130 MeV } &  \multicolumn{3}{c|}{ (0.24 GeV)$^3$ }       \\     
\multicolumn{2}{|c|}{  calc }  &   \multicolumn{3}{c|}{ 5.5 MeV}  &  \multicolumn{3}{c|}{  125 MeV }  &   \multicolumn{3}{c|}{  (0.241 GeV)$^{3\dagger}$  }       \\ \hline
        & $m_\pi$ & $f_\pi$ & $m_K$ & $f_K$   &   $m_\rho$ &  $f_\rho$  & $m_K^\star$ & $f_K^\star$ &  $m_\phi$ &  $f_\phi$
 \\ \hline 
 expt  &   0.138  &  0.131  &   0.496   &   0.160 &   0.770  &  0.216  &   0.892 &  0.225   &   1.020   &  0.236    \\  
 calc  &   0.138$^\dagger$ &  0.131$^\dagger$ & 0.497$^\dagger$ &  0.155  &  0.742 & 0.207 & 0.936 & 0.241  &  1.072     &    0.259   \\ \hline  
\end{tabular}
\end{table}

We employ the ladder-rainbow kernel found to be successful in earlier 
work for light quarks~\cite{Maris:1997tm,Maris:1999nt}.   It can be written
\mbox{$\alpha_{\rm eff}(k^2) =  \alpha^{\rm IR}(k^2) + \alpha^{\rm UV}(k^2) $}
The IR term implements the strong infrared enhancement in the region
\mbox{$0 < k^2 < 1\,{\rm GeV}^2$} required for sufficient dynamical
chiral symmetry breaking.   The UV term preserves the one-loop renormalization group behavior of QCD: \mbox{$\alpha_{\rm eff}(k^2) \to \alpha_s(k^2)^{\rm 1 loop}(k^2)$} in the ultraviolet with 
\mbox{$N_f=4$} and \mbox{$\Lambda_{\rm QCD} = 0.234\,{\rm GeV}$}.   The strength of $\alpha^{\rm IR}$ along with two quark masses are fitted to $\langle\bar q q\rangle $, $m_{\pi/K}$ and
$f_{\pi}$.  This and selected light quark vector meson results are displayed in Table~\ref{Table:model}.
The momentum dependence of the kernel along with the IR and UV components is displayed in Fig.~\ref{Fig:MT_kernel}.    The infrared component of this effective kernel is phenomenological because QCD is unsolved in such a non-perturbative domain. To help replace such phenomenology by specific mechanisms, it is necessary to first  characterize its performance in new domains.

\section{Results for Heavy Quark Mesons}

In Table~\ref{Table:qQ} we display the results for the heavy-light  ground state pseudoscalars and vectors involving a c-quark or b-quark.   We use DSE solutions for the dressed light quarks.  If a constituent mass propagator is used for the heavy quark, with the constituent mass obtained from a fit to the lightest pseudoscalar, the various meson masses are easily reproduced.   The constituent masses found this way are $M_c^{\rm cons} = 2.0 $~GeV for the c-quark, and $M_b^{\rm cons} = 5.3 $~GeV for the b-quark.   To compare with what is known about quark masses, we take the quark current mass values~\cite{PDG04} \mbox{$m_c = 1.2\pm0.2$}~GeV, and \mbox{$m_b = 4.2\pm0.2$}~GeV at scale $\mu =$ 2~GeV and use the quark DSE to run the masses into the timelike 
region where the meson mass shells are located.   If all the meson momentum runs through the heavy quark, the mass function from the DSE would suggest an effective quark mass  
$M_q^{DSE}(p^2\sim -M^2)$ where M is the meson mass.  In this domain the mass function varies slowly and reproduces the previously obtained values for $M_{c/b}^{\rm cons}$ within 10\%.  In this sense, heavy quark dressing is well summarized by a constituent mass.
However, the electroweak decay constants obtained from the constituent mass approximation are 
30-50\% below the available experimental values.    Moreover, within this ladder-rainbow model, quark dressing is not a minor effect because the use of fully dressed quark propagators, both heavy and light, does not yield a physical bound state solution for these heavy-light states involving a c-quark or b-quark.   
\begin{table}
\caption{Calculated masses and electroweak decay constants for ground state pseudoscalar 
and vector heavy-light mesons, together with experimental data~\protect\cite{PDG04}, all in GeV.   In the rows labelled {\it calc M}, the heavy quark is described by a constituent mass fit to the lightest pseudoscalar (marked by $\dagger$).    In the rows labelled {\it $\Sigma_{\rm UV}$ only}, the heavy quark is dressed, but  only by the UV component of the DSE kernel; this did not produce physical solutions for $D$ or $D^*$ states.
\label{Table:qQ} }

\begin{tabular}{|l|cc|cc|cc|cc|cc|} \hline 
      & D & D$^*$ & D$_s$ & D$^*_s$   &   B &  B$^{*}$ & B$_s$ & B$^{*}_s$ &  B$_c$ &  B$^{*}_c$
 \\ \hline 
 expt M &   1.86  &  2.01        &   1.97    &   2.11 &   5.28  &  5.33  &   5.37 &  5.41   &   6.29   &     ?    \\  
 calc M &   1.85$^\dagger$ &  2.04 & 1.97 &  2.17  &  5.27$^\dagger$ & 5.32 & 5.38 & 5.42  &   6.36     &    6.44   \\ 
 $\Sigma_{\rm UV}$ only &  - & -  & -  & -   &   4.66  &     & 4.75   &    &  5.83  &    \\ \hline  
 expt f   &  0.222                  &      ?     & 0.294 &      ?   &  0.176                  &       ?     &    ?    &    ?    &      ?       &        ?    \\ 
 calc f    & 0.154                  &  0.160 &0.197 & 0.180       & 0.105                  &  0.182 & 0.144 & 0.20 &  0.210   &  0.18      \\ 
 $\Sigma_{\rm UV}$ only &  - & -  & -  & -   &   0.133  &     &  0.164   &    &  0.453  &    \\ \hline  
            
\end{tabular}
\end{table}

The results for equal quark mesons (quarkonia), displayed in Table~\ref{Table:quarkonia}, show a different perspective.    With the same fitted constituent masses, the meson masses are well reproduced, and again the electroweak decay constants are  too low by some 30-70 \%.    Use of dynamically dressed propagators removes almost all of this deficiency and the decay constants are well reproduced. This improvement provided by dynamical dressing of c- and b-quarks is persistent and systematic in the following sense. When the dressing is  progressively  introduced into all three stages (bound state solution, normalization loop integral, and then the loop integral for evaluation of the decay constant), the result always increases towards the experimental value. This suggests that a constituent mass approximation, even for b-quarks, is inadequate. Small departures from a strictly constant mass function and field renormalization function $Z(p^2)$  for quarks in the relevant region of the complex plane are magnified due to the very weak binding of the mesons in question. 

With increasing quark mass, the quarkonia states become smaller in size and the ultraviolet sector of the  ladder-rainbow kernel becomes more influential.   The size of the dressed quark quasi-particle also decreases with quark mass so that the entire heavy quarkonia dynamics should be dominated by the ultraviolet  sector of the kernel as dictated by pQCD.    In the last two rows of Table~\ref{Table:quarkonia}, we display the results obtained by using just the infrared sector, or just the ultraviolet sector, of the ladder-rainbow kernel in all phases of the calculation, binding and quark dressing.     For ease of calculation, we define these two components  of the kernel as shown in Fig.~\ref{Fig:MT_kernel}, corresponding to the two terms in the kernel expression~\cite{Maris:1997tm,Maris:1999nt}.   With only the UV kernel, a mass shell couldn't be reached for $c \bar c$,  but excellent masses for the $b \bar b$ states are obtained.
\begin{table}
\caption{Calculated masses and electroweak decay constants for ground state pseudoscalar 
and vector quarkonia, together with experimental data~\protect\cite{PDG04}, all in GeV.  Results employ either a constituent mass treatment of the heavy quark, labelled {\it $ M_Q^{\rm cons}$}, or a dynamical dressed propagator, labelled {\it $ \Sigma_Q^{\rm  DSE}(p^2)$}.    The last two rows shown the \% change from the full dynamical calculation due to retention of just the IR or the UV component of the ladder-rainbow kernel in all aspects of the calculation, both quark dressing and binding.
\label{Table:quarkonia} }

\begin{tabular}{|l|cc|cc|cc|cc|} \hline 
         & $M_{\eta_c}$  &  $f_{\eta_c}$   & $M_{J/\psi}$  &   $f_{J/\psi}$    & $M_{\eta_b}$  &  $f_{\eta_b}$   & $M_\Upsilon$  &   $f_\Upsilon$     \\ \hline 
 expt                    &   2.98                &     0.340             &      3.09           &  0.411        &   9.4 ?                &            ?           &      9.46           &  0.708              \\ 
 calc  with $ M_Q^{\rm cons}$ &3.02&    0.239          &      3.19           &  0.198    &9.6&    0.244          &      9.65           &  0.210              \\ 
 calc  with $ \Sigma_Q^{\rm  DSE}(p^2)$&3.04&    0.387   &     3.24           &   0.415   &9.59&  0.692   &     9.66           &   0.682    \\   \hline
IR only    &   -21\%    &    -29\%   &    -20\%    &    -20\%    &   -15\%    &    -75\%    &    -16\%    &   -50\% \\ 
UV only   &   -             &      -           &       -          &      -           &  -0.6\%    &     -12\%  &   -0.8\%     &  -25\%  \\   
\hline  
\end{tabular}
\end{table}

In the case of heavy-light mesons,  the binding and dressing effect for the light quark is not significantly different  from the dynamics of light quarkonia and the empirical chiral condensate  that set the characteristic infrared length scale and strength of the employed ladder-rainbow kernel.    For this reason we investigated a hybrid procedure in which the ultraviolet component of the kernel was used for heavy quark dressing, while the full kernel was used for binding and light quark dressing.    The results displayed in  Table~\ref{Table:qQ} show that physical $B$ states are produced this way, whereas that was impossible with a b-quark dressed via the full kernel.    This again suggests that for heavier mass, or smaller size objects, the infrared sector of the present kernel is too strong.    In the case of heavy quarkonia, the bound state wavefunction itself provides some limitation of support in the BSE integrand to small distance.    However that is less the case for the heavy-light states;  a forced suppression of the infrared sector of the heavy quark DSE seems necessary.

This may indicate the present kernel, determined for light quarks, has implicit infrared strength through quark-gluon vertex dressing that should be much reduced for heavy quarks.   
Aspects of this study may also be  a partial  confirmation of Brodsky and Shrock's suggestion of a universal maximum wavelength of quarks and gluons in hadrons~\cite{Brodsky:2008be}.   The resulting minimum momentum would increase with mass of the hadron or dressed quark/gluon constituent and such a cutoff may systematically link our findings.   

\section{Leading npQCD scale and the four quark condensate}

Quark helicity and chirality in QCD are increasingly good quantum numbers at short distances or at momentum scales significantly larger than any mass scale.   One manifestation of this is that, for chiral quarks, the correlator of a pair of vector currents is identical to the corresponding correlator of a pair of axial vector currents to all finite orders of pQCD.   In the non-perturbative circumstance, the difference of such correlators measures chirality flips, and the leading non-zero contribution in the ultraviolet identifies the leading non-perturbative phenomenon in QCD.     This is the four quark 
condensate~\cite{Narison:1989aq}.    The ladder-rainbow kernel can produce the difference correlator as a vacuum polarization integral in momentum space, where the propagators are dressed and the vector and axial vector vertices are generated in a way consistent with the symmetries.   We use the large spacelike momentum dependence to extract the leading coefficient or condensate; and a Fourier transform identifies the leading non-perturbative distance scale.
\begin{figure}[ht]
\includegraphics[height=0.28\textheight]{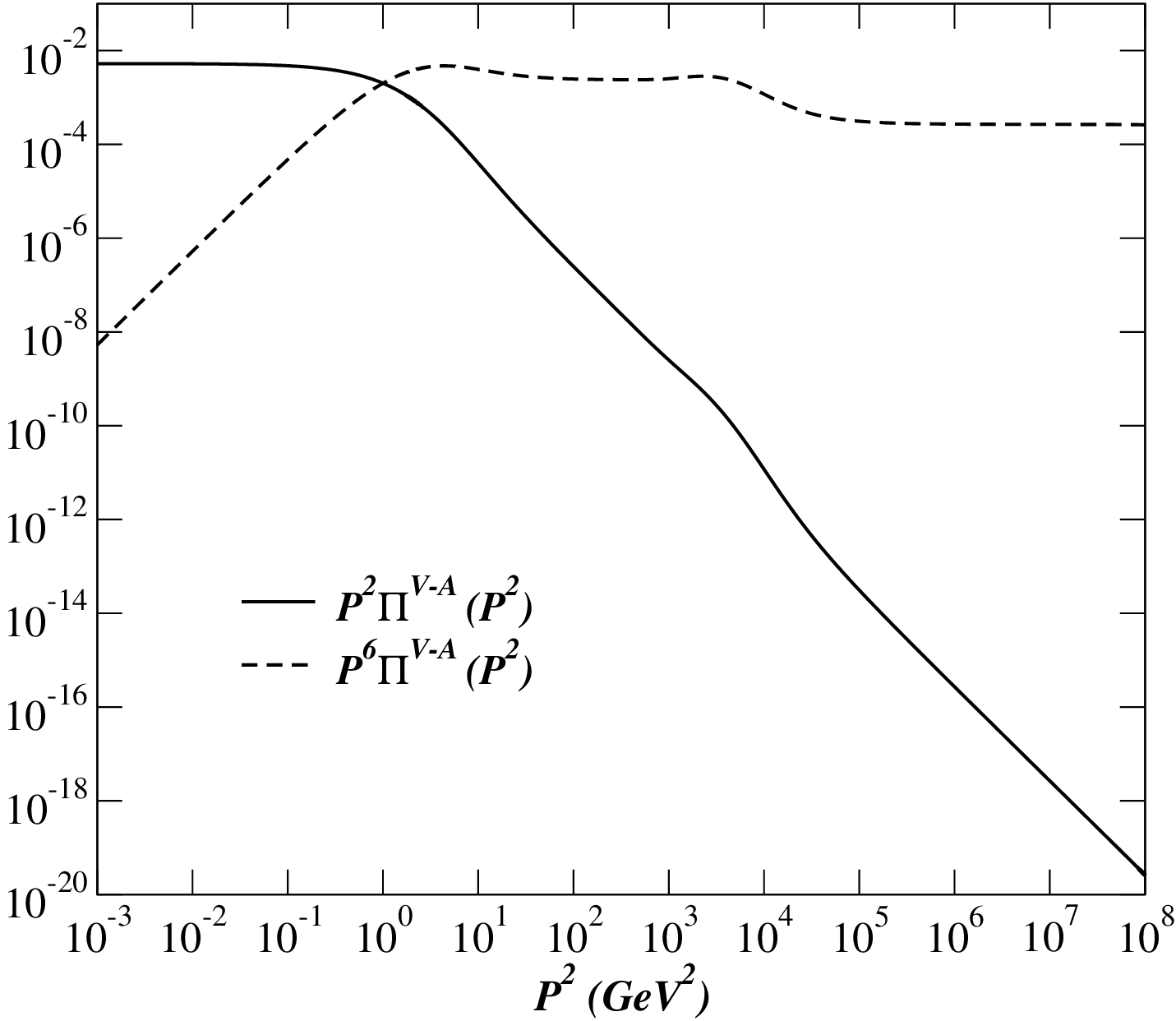}
\includegraphics[height=0.28\textheight]{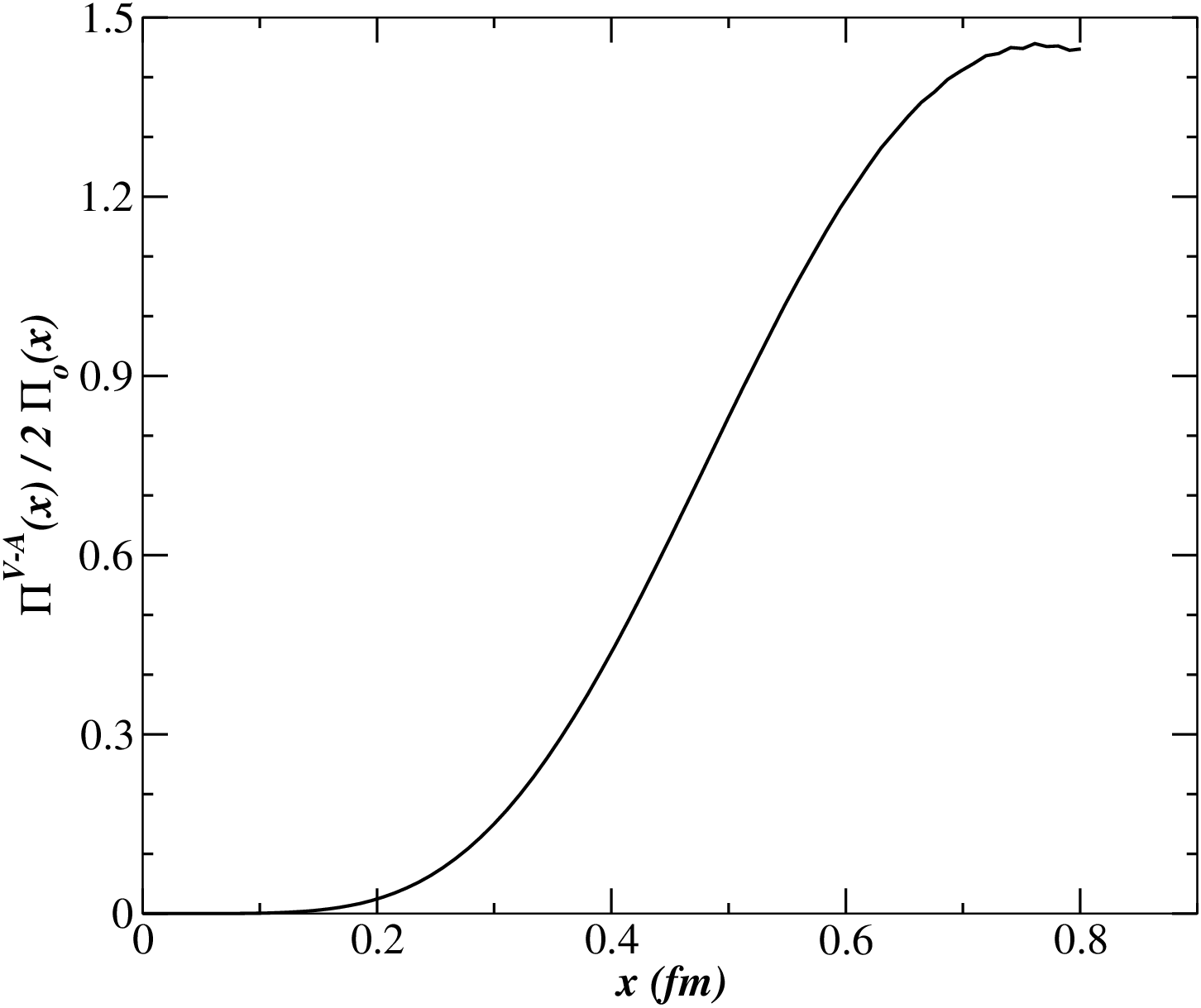}
\caption{{\it Left Panel}: $P^n\, \Pi_{T}^{V-A}(P^{2})$ for \mbox{$n =  2, 6$};  the latter allows extraction of 
$\langle \bar{q}q\bar{q}q\rangle$.     {\it Right Panel}:  The ratio of   $\Pi_{T}^{V-A}(x)$ to the free summed correlator $2 \Pi_{T}^0(x)$. \label{fig:corratiopD} }
\end{figure}

The vector current-current correlator is formulated as the loop integral
\begin{equation}
\Pi_{\mu\nu}^{V}(P) = \int^\Lambda_q d^{4}x \; {\rm e}^{iP \cdot x} 
\langle 0|T \, j_\mu(x)\, j^+_\nu(0)|0 \rangle = -\int \frac{d^{4}q}{(2\pi)^4} 
Tr\lbrace \gamma_\mu S(q_+) \Gamma^{V}_{\nu} (q,P)S(q_-)\rbrace ~~,
\label{Eq:vcorr}
\end{equation}
where $\Lambda $ indicates regularization, e.g., by the Pauli-Villars method, and $\Gamma^{V}_{\nu}$ is the dressed vector vertex.   The axial vector correlator is formulated in an analogous way and we directly calculate the difference correlator which does not require ultraviolet regularization.   
With \mbox{$\Pi_{\mu\nu}^{V}(P) = (P^{2}\delta_{\mu\nu} - P_{\mu} P_{\nu})\Pi_{T}^{V}(P^{2})$}, and 
\mbox{$\Pi_{\mu\nu}^{A}(P) = (P^{2} \delta_{\mu\nu} - P_{\mu} P_{\nu})
\Pi_{T}^{A}(P^{2})+ P_{\mu} P_{\nu}\, \Pi_{L}^{A}(P^{2}) $}, the quantity of interest here is 
\mbox{$\Pi_{T}^{V-A}(P^{2}) = \Pi_{T}^{V}(P^{2}) - \Pi_{T}^{A}(P^{2}) $}.  

The leading non-perturbative contribution  to $\Pi_{T}^{V-A}$ starts with 
dimension $d =6$ and involves the four-quark condensate in the form~\cite{Dominguez:1998wy,Dominguez:2003dr}
\begin{equation}
\label{eq:c4cm}
\Pi_{T}^{V-A}(P^{2}) = -\frac{32 \pi}{9} 
\frac{\alpha_{s} \langle \bar{q}q\bar{q}q\rangle }{P^{6}}
\{1+\frac{\alpha_{s}(P^{2})}{4\pi}[\frac{247}{12}+ {\rm ln} (\frac{\mu^{2}}{P^{2}})]
\} + O(\frac{1}{P^{8}})~~.
\end{equation}
The left panel of Fig.~\ref{fig:corratiopD} indicates that our numerical calculation of $P^6\, \Pi_{T}^{V-A}(P^{2})$ identifies a leading ultraviolet constant reasonably well.   The four quark condensate 
$\langle \bar{q}q\bar{q}q\rangle $ extracted via Eq.~(\ref{eq:c4cm}) is 65\% greater than the common vacuum saturation assumption $\langle \bar{q}q\rangle^2$ at the renormalization scale \mbox{$\mu = 19$}~GeV used in this work.  
The low $P^2$ limit provides a reasonable account of the first Weinberg sum rule~\cite{Weinberg:1967kj, Dorokhov:2003kf}:   
\mbox{$P^{2} \,\Pi_{T}^{V-A}(P^{2})|_{P^{2} \to 0}  = - f_{\pi}^{2} $}, in the \mbox{$f_\pi = 0.0924$} GeV convention.    We obtain  \mbox{$f_\pi = 0.0728$} GeV that way.   Our results are consistent with the  second Weinberg sum rule~\cite{Weinberg:1967kj} \mbox{$P^4 \,\Pi_{T}^{V-A}(P^{2})|_{P^{2} \to \infty}  =0$}.    The Das-Guralnik-Mathur-Low-Young sum rule~\cite{Das:1967it} relates 
\mbox{$\int_{0}^{\infty}{dP^{2}}\, P^{2}\,\Pi_{T}^{V-A}(P^{2})$} to  the electromagnetic component of $m_{\pi^{\pm}} - m_{\pi^{0}}$.   We obtain 4.86 MeV for this mass difference in comparison with $4.43 \pm 0.03$  from experiment.   

The right panel of Fig.~\ref{fig:corratiopD} displays the x-dependence of the V-A correlator amplitude expressed as a ratio to the free V$+$A correlator amplitude.    The latter  diverges in the ultraviolet 
as ${\rm ln}(P^2)$, which corresponds to $ x^{-4}$ at small $x$.   This chirality-flip ratio identifies a scale of $\sim 0.5$~fm for the onset of non-perturbative dynamics.   This ladder-rainbow truncation within a DSE format produces the same scale as obtained from the ratio of P-S and P$+$S correlators in both a lattice-QCD calculation and the Instanton Liquid Model~\cite{Faccioli:2003qz}.   


\begin{theacknowledgments}
The authors would like to thank P. Maris, C. D. Roberts and S. J. Brodsky for helpful conversations
and suggestions.   This work has been partially supported by  the U.S. National Science Foundation under grant no. \ PHY-0610129.
\end{theacknowledgments}
\bibliography{SanCarlos08_proc}


\bibliographystyle{aipproc}   


\IfFileExists{\jobname.bbl}{}
 {\typeout{}
  \typeout{******************************************}
  \typeout{** Please run "bibtex \jobname" to optain}
  \typeout{** the bibliography and then re-run LaTeX}
  \typeout{** twice to fix the references!}
  \typeout{******************************************}
  \typeout{}
 }

\end{document}